\documentclass[preprint,showpacs,preprintnumbers,amsmath,amssymb]{revtex4}

\usepackage{graphicx}
\usepackage{dcolumn}
\usepackage{bm}

\begin{document}


\title{Study of elliptical flow at VECC-SCC500 energies   
\\}

\author{Varinderjit Kaur}
\email{vkaur@thapar.edu}
\author{Suneel Kumar}%

\affiliation{%
School of Physics and Materials Science, Thapar University Patiala-147004, Punjab (India)\\
}
\author{Rajeev K. Puri}%
\affiliation{
Department of Physics, Panjab University, Chandigarh (India)\\
}%
\author{S. Bhattacharya}%
\affiliation{
Variable Energy Cyclotron Center, 1/AF, Bidhan Nagar, Kolkata INDIA\\
}%

\date{\today}
\maketitle
\baselineskip=18pt

\section{Introduction}
Upcoming facility of Super Conducting Cyclotron (SCC500) at Variable Energy
Cyclotron Center (VECC) Kolkata will give scientists an opportunity to 
study the various phenomena at intermediate energies. 
The behavior of nuclear matter under the extreme conditions of temperature, density, 
angular momentum etc., is a very important aspect of heavy-ion physics. One of the important quantity 
which has been used extensively to study this hot and dense nuclear matter is the elliptical 
flow.
Recently, the analysis of the transverse-momentum dependence of elliptical flow 
has also been put forwarded \cite{lukasik}. 
The elliptical
flow is shaped by the interplay 
between the geometry and the mean field 
and, 
when gated by the transverse momentum, 
reveals the momentum dependence of the mean field at supra-normal densities.
The elliptical flow describes the eccentricity of an ellipse like distribution.
Quantitatively, it is the difference between the major and minor axis.
The orientation of the major axis is confined to azimuthal angle
$\phi$ or $\phi$+$\frac{\pi}{2}$ for
ellipse like distribution. The major axis lies within the reaction plane
for $\phi$; while $\phi$+$\frac{\pi}{2}$
indicates that the orientation of the ellipse is perpendicular to the reaction plane, which is the case for
squeeze out flow and may be expected at mid rapidity \cite{voloshin}.
The elliptical flow 
is quantified by the second-order Fourier coefficient \cite{voloshin}
\begin{equation}
V_2~=~ \langle \cos2\phi \rangle=\left\langle\frac{p_x^2 - p_y^2}{p_x^2 + p_y^2}\right\rangle,
\end{equation}
from the azimuthal distribution of the
detected particles at mid rapidity as:
\begin{equation}
\frac{dN}{d\phi} = p_0(1 + 2V_1\cos\phi + 2V_2\cos2\phi+\ldots),
\end{equation}
where $p_x$ and $p_y$ are the $x$ and $y$ components of  the momentum. 
The $p_x$ is in the reaction plane, 
while $p_y$ is perpendicular to the reaction plane,
and $\phi$ is the azimuthal angle of the emitted particle's momentum 
relative to the $x$-axis. 
The positive values of $\langle\cos 2\phi\rangle$
reflect preferential in-plane emission, 
while negative values reflect preferential out-of-plane emission. 
\section{The Model}
The model is the semi classical microscopic improved version of QMD model \cite{hartnack} 
where nucleons interact via two and three body interactions. The nucleons propagate
according to classical Hamilton equations of motion. 
\begin{equation}
\frac{d\vec{r_i}}{dt} = \frac{d<H>}{d\vec{p_i}}~~~~;~~~~\frac{d\vec{p_i}}{dt} = -\frac{d<H>}{d\vec{r_i}}.
\end{equation}
 with\\ 
${ <H> = <T> + <V>}$  is the Hamiltonian.
The total interaction potential is given as: 
\begin{eqnarray}
V^{ij}(\vec{r}^\prime -\vec{r})&=&V^{ij}_{Skyrme}+V^{ij}_{Yukawa}+V^{ij}_{Coul}+\nonumber\\
& &V^{ij}_{mdi}+V^{ij}_{sym}
\end{eqnarray}\\
where ${V_{Skyrme}}$,  ${V_{Yukawa}}$,  ${V_{Coul}}$, ${V_{mdi}}$ and ${V_{sym}}$ represent, 
respectively, the Skyrme, Yukawa, Coulomb, momentum-dependent interaction (MDI) and symmetry
potentials.
\section{Results and Discussion}
For the controlled study of the role of mass asymmetry of a
reaction \cite{plb}, we simulated several thousands events of various
reactions by keeping the total reacting mass (= 152 units). While
the total mass stays constant, mass asymmetry parameter ${\eta}$ is varied
by choosing different combinations of projectiles-targets. These projectile-target
combinations are possible at SCC-500 developed at VECC. We have
performed exclusive studies of elliptical flow by simulating the reactions of
$_{24}Cr^{50}+_{44}Ru^{102}$ (${\eta = 0.3}$),
$_{16}S^{32}+_{50}Sn^{120}$ (${\eta = 0.5}$), and
$_{8}O^{16}+_{54}Xe^{136}$ (${\eta = 0.7}$) at incident energies
between 50 MeV/nucleon for semi-central impact parameter
using a soft nuclear equation of state. \\
\begin{figure}
\includegraphics{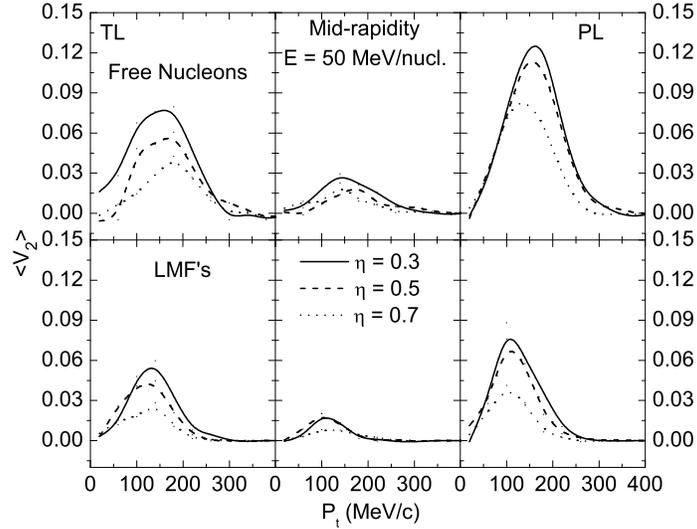}
\caption{\label{fig:1} The transverse momentum dependence of the elliptical flow at E =
50 MeV/nucleon for different mass asymmetries divided into contributions from target-like,
mid-rapidity and projectile-like matter, respectively. The upper and lower panels represent the 
free nucleons 
and light mass fragments (LMF's), respectively.}  
\end{figure}
Fig. 1 shows the elliptical flow for the
free particles (A=1)(upper panel) and light mass fragments (LMF's)
[${(2\le A \le4)}$] (lower panel) as a function of transverse
momentum ${(P_t)}$ at an incident beam energy E = 50 MeV/nucl.
The different curves in each
panel correspond to different mass asymmetries.
Moreover, we divide the total elliptical flow into contributions from target-like(TL) (left panels),
mid-rapidity (middle panels),
and projectile-like(PL) (right panels) particles.
From the figure, we see that the projectile-like (PL) nucleons
and LMF's feel more squeeze out compared to target-like (TL) nucleons/LMF's. At the larger
mass asymmetry, only small fraction of nucleons/LMF's experience squeeze out compared to symmetric
reactions. This decrease of squeeze out with mass asymmetry happens due to decreasing participant zone.
This is in agreement with earlier calculations where fragments were found to exhibit similar trends.
The detailed study in this direction will be fruitful for experimentlists. 

\end{document}